\pdfoutput=1

\documentclass[11pt]{article}

\usepackage{EMNLP2023}

\usepackage{times}
\usepackage{latexsym}
\usepackage{balance}
\usepackage[T1]{fontenc}

\usepackage[utf8]{inputenc}

\usepackage{microtype}

\usepackage{inconsolata}
\usepackage{graphicx}

\title{RETA-LLM: A Retrieval-Augmented Large Language Model Toolkit}

\author{Jiongnan Liu$^{1}$, Jiajie Jin$^{2}$, Zihan Wang$^{1}$, Jiehan Cheng$^{1}$, Zhicheng Dou$^{1*}$, \and Ji-Rong Wen$^{1}$ \\
  $^{1}$Gaoling School of Artificial Intelligence, Renmin University of China \\
  $^{2}$University of Science and Technology of China \\
  \texttt{$^{1}$\{liujn, wangzihan0527, jiehan\_cheng, dou, jrwen\}@ruc.edu.cn} \\ \texttt{$^{2}$jinjiajie@mail.ustc.edu.cn}}

\begin{document}
\maketitle

\def\thefootnote{*}\footnotetext{Corresponding author.}
\def\thefootnote{\arabic{footnote}}

\begin{abstract}
Although Large Language Models (LLMs) have demonstrated extraordinary capabilities in many domains, they still have a tendency to hallucinate and generate fictitious responses to user requests. This problem can be alleviated by augmenting LLMs with information retrieval (IR) systems (also known as retrieval-augmented LLMs). Applying this strategy, LLMs can generate more factual texts in response to user input according to the relevant content retrieved by IR systems from external corpora as references. In addition, by incorporating external knowledge, retrieval-augmented LLMs can answer in-domain questions that cannot be answered by solely relying on the world knowledge stored in parameters. To support research in this area and facilitate the development of retrieval-augmented LLM systems, we develop RETA-LLM, a {RET}reival-{A}ugmented LLM toolkit. In RETA-LLM, we create a complete pipeline to help researchers and users build their customized in-domain LLM-based systems. Compared with previous retrieval-augmented LLM systems, RETA-LLM provides more plug-and-play modules to support better interaction between IR systems and LLMs, including {request rewriting, document retrieval, passage extraction, answer generation, and fact checking} modules.  Our toolkit is publicly available at {\url{https://github.com/RUC-GSAI/YuLan-IR/tree/main/RETA-LLM}}.

\end{abstract}

\section{Introduction}
\label{sec:intro}

Large language models (LLMs) have attracted increasing attention from both research community and industry~\cite{GPT3, GPT4, instructGPT, LLAMA, PALM, LLMsurvey, chatglm1}. With tremendous world knowledge stored in parameters~\cite{world-knowledge-1, world-knowledge-2, world-knowledge-3} and the Reinforcement Learning from Human Feedback (RLHF) techniques~\cite{RLHF, RLHF-2}, LLMs can generate helpful, detailed, and polite texts in response to user inputs. Many studies have demonstrated LLMs' extraordinary abilities in various areas, including nature language processing~\cite{LLMnlp}, information retrieval~\cite{searchGPT, Query2Doc, LLMCDR}, and recommendation~\cite{LLMREC, LLMREC-2}.

However, LLMs still tend to hallucinate and sometimes generate texts opposite to facts~\cite{hallucinate, LLMsurvey}. To tackle these problems, researchers have proposed a new paradigm to strengthen LLMs with information retrieval systems (retrieval-augmented LLMs)~\cite{RAG1, RAG2, WebGPT}, which enables LLMs to retrieve relevant contents from an external repository (knowledge corpus) to generate texts based on them. It has been verified that retrieval-augmented LLMs can generate texts in response to user input with fewer hallucinations~\cite{WebGPT}. Furthermore, by incorporating customized private data resources, retrieval-augmented LLMs can respond to in-domain queries that cannot be answered by LLMs trained with public data.

\begin{figure*}[!tbp]
\setlength{\abovecaptionskip}{0.1cm}
\setlength{\belowcaptionskip}{0.1cm}
\centering
\includegraphics[width=\linewidth]{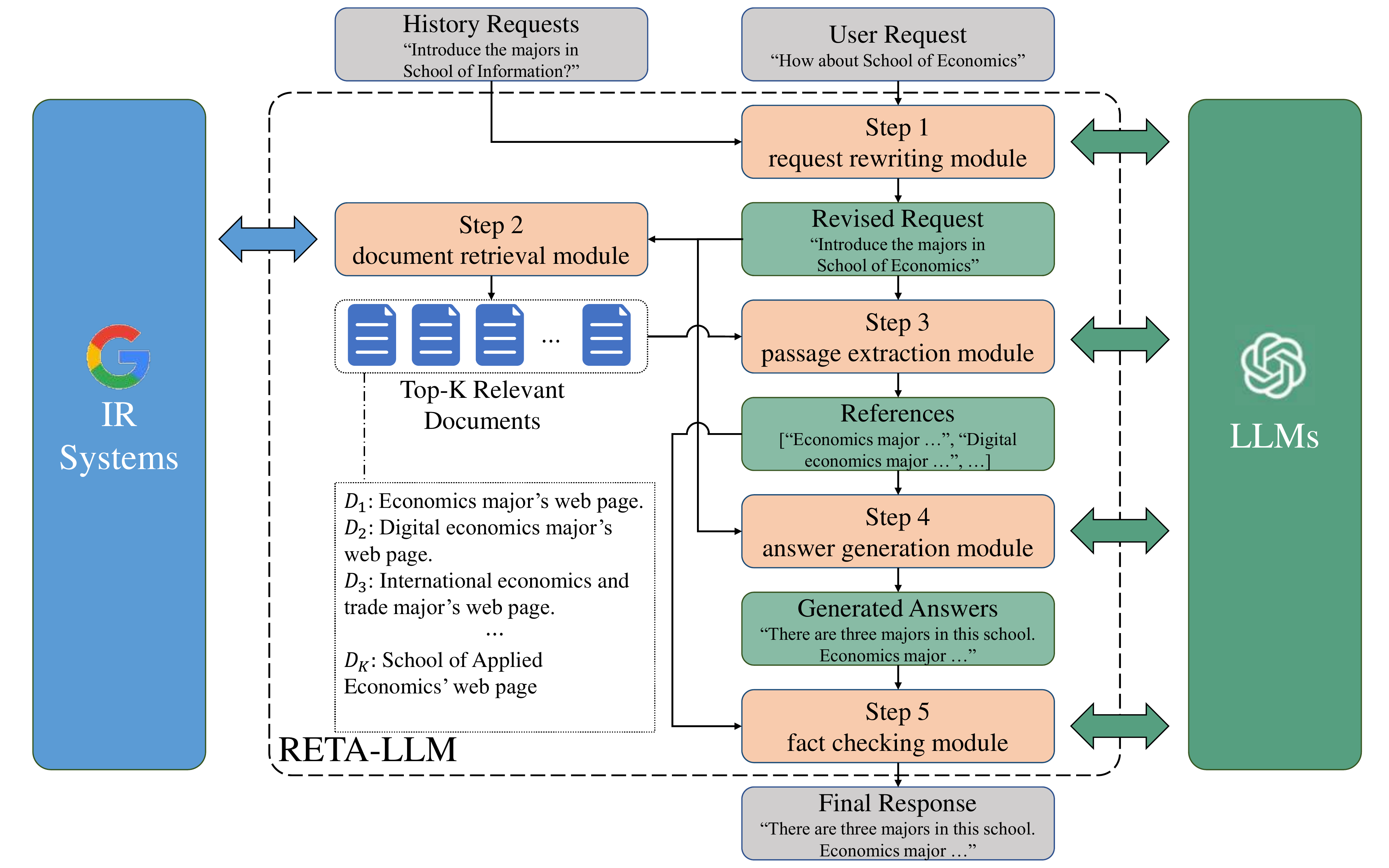}
\caption{The RETA-LLM framework. Examples are taken from an intelligent university information seeking system powered by RETA-LLM.}
\label{fig:framework}  
\end{figure*}

To support research in this area and help users build their own in-domain LLM-based systems, we devise RETA-LLM, a \textbf{RET}reival-\textbf{A}ugmented LLM toolkit. Different from previous general LLM-enhanced toolkits such as LangChain,\footnote{LangChain, \url{https://github.com/hwchase17/langchain}} RETA-LLM focuses on the retrieval-augmented LLMs and provides more plug-in modules. Typically, retrieval-augmented LLMs use a retrieve-and-generate strategy with two modules: First, they retrieve documents or passages based on user request (\textbf{document retrieval} module); then, they generate answers utilizing these relevant documents as references (\textbf{answer generation} module). In addition to these two basic modules, our RETA-LLM  provides three optional modules: (1) a \textbf{request rewriting} module to make user's current request more complete and clear; (2) a \textbf{passage extraction} module to extract relevant passages or fragments from the whole retrieved document contents; and (3) a \textbf{fact checking} module to verify whether there exist factual errors in the generated answers. These optional modules can make the interaction between IR systems and LLMs more effective and smooth. The disentanglement between LLMs and IR systems in our RETA-LLM is more thorough, which makes the customization of search engines and LLMs more convenient. Furthermore, to make the usage easier, we provide a complete and ready-to-use pipeline for researchers and users to build their RETA-LLM toolkits based on their own repository for in-domain LLM-based systems from scratch.

RETA-LLM is part of YuLan, a open source LLM initiative proposed by Gaoling School of Artificial Intelligence, Renmin University of China. RETA-LLM is still under development and there are many issues that need to be solved with great efforts. We sincerely welcome contributions on this open source toolkit.

\section{RETA-LLM Framework}
As aforementioned, compared with Langchain, which is a common LLM-augmented toolkit, our RETA-LLM toolkit focuses specifically on retrieval-augmented LLMs. We provide five plug-in modules in RETA-LLM to interact with LLMs and IR systems. The modules include request rewriting, document retrieval, passage extraction, answer generation, and fact checking modules. The framework of our RETA-LLM is shown in Figure~\ref{fig:framework}. The workflow of RETA-LLM is as follows:

{First, RETA-LLM uses the request rewriting module to revise the current user request to make it complete and clear.} Because users can issue a series of questions to the RETA-LLM, the semantics of the current user request may be incomplete. For example, A user may ask \textit{``How about the School of Economics?''} while the historical request is \textit{``Introduce the majors in School of Information''}. In this case, the precise meaning of the user is \textit{``Introduce the majors in School of Economics''}. Since LLMs have shown remarkable abilities in rewriting queries in conversational dense retrieval~\cite{LLMCDR}, we feed the current user request and the previous conversation histories to LLMs to perform rewriting.

{Then, RETA-LLM uses the document retrieval module to retrieve relevant documents from the external corpus based on the revised user request.} The document retrieval module is the module connected to the IR system. It retrieves relevant documents from the external knowledge corpus and returns top-$K$ of them. The $K$ is set to 3 in our default configuration. We provide a default dense retriever in our repository. The detailed description can be found in the next section.

{Next, RETA-LLM uses the passage extraction module to extract fragments related to the user request from the retrieved documents to form the references.} Because of the input length limitations (typically 2048 or 4096 tokens) of LLMs, it is impossible to directly concatenate the contents of all top-$K$ relevant document contents as references for them to generate answers. Trivial methods by truncating the document contents may lose important information in them. Therefore, we reuse the LLMs themselves to extract related fragments from retrieved documents based on the revised request. Since the length of one document may also exceed the limitations, we apply the sliding window strategy to extract fragments step by step. The sliding window size and step are set to 512 and 256 in our default configuration. These fragments are then concatenated together as the references.

{Besides, RETA-LLM uses the answer generation module to generate answers for the user request.} As previous researches~\cite{WebGPT, RAG1, RAG2} suggest, by feeding the references retrieved from the external corpus, LLMs can generate more factual answers.

{Finally, RETA-LLM uses the fact checking module to verify whether the generated answers contain factual mistakes and output final responses for the user request.} Though providing additional evidence for generation, LLMs may also hallucinate~\cite{WebGPT}. It is necessary to devise a module to conduct further fact verification. Because of the strong natural language understanding abilities of LLMs, we feed the references and generated answers to them to make judgments. Therefore, RETA-LLM can decide whether to output the generated answers or just say ``\textit{I cannot answer this question}''.

Noticed that all the inputs to the LLMs are wrapped in instructions or prompts. As shown in Figure~\ref{fig:framework}, we disentangle the IR systems and LLMs entirely in our RETA-LLM. This separate design in our RETA-LLM leads users can customize their personal search engines and LLMs.


\section{RETA-LLM Usage Pipeline}
To make the toolkit more convenient for personal usage, we provide a complete pipeline to build in-domain LLM-based system based on html resources. The pipeline is as follows:

{First}, RETA-LLM uses \texttt{Beautiful Soup} package to convert the raw html files into json data in our \textbf{HTML Converter}.\footnote{Beautiful Soup, \url{https://beautiful-soup-4.readthedocs.io/en/latest/}}

{Second}, RETA-LLM follows the implementation of \texttt{disentangled-retriever}~\cite{disentangled-retriever} to build dense indexes and to conduct domain adaption from the converted json data in our \textbf{Index Builder}.\footnote{disentagled-retriever, \url{https://github.com/jingtaozhan/disentangled-retriever}} Specifically, our method supports unsupervised training of dense retrieval models on local document collections, enabling the model to learn domain-specific knowledge in advance. Compared with the retrieval module in the popular LangChain library, our retrieval method has two advantages: (1) the model learns knowledge within the domain of local documents, enabling it to match queries more accurately, and (2) our method does not segment text, thus avoiding any negative impact on the overall semantic information of the text. We also provide a sparse retriever applying \texttt{faiss}~\cite{faiss} package to build sparse indexes.\footnote{Faiss, \url{https://github.com/facebookresearch/faiss}}
Otherwise, users can also use their customized search engines as the document retrieval module.

{Third}, users need to prepare LLMs for question answering. For LLM loading and responding, we provide the template for  Alpaca~\cite{alpaca},\footnote{Alpaca,\url{https://github.com/tatsu-lab/stanford_alpaca}}, YuLan-Chat,\footnote{YuLan-Chat, \url{https://github.com/RUC-GSAI/YuLan-Chat}} ChatGLM~\cite{chatglm1, chatglm2},\footnote{ChatGLM, \url{https://github.com/THUDM/ChatGLM-6B}} and GPT-3.5 API~\cite{instructGPT}.\footnote{OpenAI's API, \url{https://api.openai.com/v1/completions}} If users use other LLMs, they can edit the codes and configurations in our toolkit.

{Finally}, users can start their own RETA-LLM services using ~\texttt{streamlit} package.\footnote{streamlit, \url{https://github.com/streamlit/streamlit}}

More details about the usage pipeline can be found on our GitHub repository.

\begin{figure}[!t]
 \setlength{\abovecaptionskip}{0.1cm}
 \setlength{\belowcaptionskip}{0.1cm}
  \centering
  \includegraphics[width=1.0\linewidth]{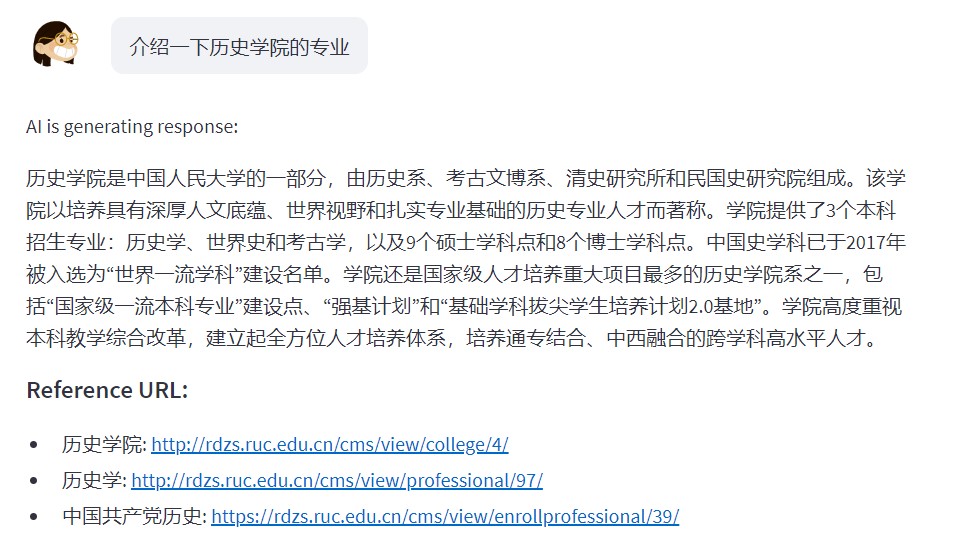}
  \caption{A case in RUC-enrollment-assistant system.}
  \label{fig:case}
\end{figure}

\section{A RETA-LLM Service Case}
Based on the RETA-LLM and the usage pipeline, we use the web pages on Renmin University of China's enrollment online platform, \footnote{Renmin University of China's enrollment online platform, \url{https://rdzs.ruc.edu.cn}} to build an RUC-enrollment-assistant system. The system uses a dense document retrieval module and adopts YuLan-13B as the backbone LLM. A using case is shown in~\ref{fig:case}. By enhancing the IR systems, LLMs can answer in-domain questions which cannot be answered by their own knowledge.

\section{Conclusion and Future Work}
In this paper, we propose RETA-LLM to facilitate research and development of retrieval-augmented LLMs. We provide five independent modules: request rewriting, document retrieval, passage extraction, answer generation, and fact checking modules in our toolkit. Furthermore, we provide a pipeline to help users build their in-domain LLM-based systems. In the future, we are going to include more retrieval-augmented LLM strategies such as active retrieval augmented generation~\cite{RAG2}. Besides, we plan to make RETA-LLM more modulized and configurable.


\newpage
\balance
\bibliography{main}

\begin{thebibliography}{26}
\expandafter\ifx\csname natexlab\endcsname\relax\def\natexlab#1{#1}\fi

\bibitem[{Brown et~al.(2020)Brown, Mann, Ryder, Subbiah, Kaplan, Dhariwal,
  Neelakantan, Shyam, Sastry, Askell, Agarwal, Herbert-Voss, Krueger, Henighan,
  Child, Ramesh, Ziegler, Wu, Winter, Hesse, Chen, Sigler, Litwin, Gray, Chess,
  Clark, Berner, McCandlish, Radford, Sutskever, and Amodei}]{GPT3}
Tom Brown, Benjamin Mann, Nick Ryder, Melanie Subbiah, Jared~D Kaplan, Prafulla
  Dhariwal, Arvind Neelakantan, Pranav Shyam, Girish Sastry, Amanda Askell,
  Sandhini Agarwal, Ariel Herbert-Voss, Gretchen Krueger, Tom Henighan, Rewon
  Child, Aditya Ramesh, Daniel Ziegler, Jeffrey Wu, Clemens Winter, Chris
  Hesse, Mark Chen, Eric Sigler, Mateusz Litwin, Scott Gray, Benjamin Chess,
  Jack Clark, Christopher Berner, Sam McCandlish, Alec Radford, Ilya Sutskever,
  and Dario Amodei. 2020.
\newblock \href
  {https://proceedings.neurips.cc/paper_files/paper/2020/file/1457c0d6bfcb4967418bfb8ac142f64a-Paper.pdf}
  {Language models are few-shot learners}.
\newblock In \emph{Advances in Neural Information Processing Systems},
  volume~33, pages 1877--1901. Curran Associates, Inc.

\bibitem[{Chowdhery et~al.(2022)Chowdhery, Narang, Devlin, Bosma, Mishra,
  Roberts, Barham, Chung, Sutton, Gehrmann, Schuh, Shi, Tsvyashchenko, Maynez,
  Rao, Barnes, Tay, Shazeer, Prabhakaran, Reif, Du, Hutchinson, Pope, Bradbury,
  Austin, Isard, Gur{-}Ari, Yin, Duke, Levskaya, Ghemawat, Dev, Michalewski,
  Garcia, Misra, Robinson, Fedus, Zhou, Ippolito, Luan, Lim, Zoph, Spiridonov,
  Sepassi, Dohan, Agrawal, Omernick, Dai, Pillai, Pellat, Lewkowycz, Moreira,
  Child, Polozov, Lee, Zhou, Wang, Saeta, Diaz, Firat, Catasta, Wei,
  Meier{-}Hellstern, Eck, Dean, Petrov, and Fiedel}]{PALM}
Aakanksha Chowdhery, Sharan Narang, Jacob Devlin, Maarten Bosma, Gaurav Mishra,
  Adam Roberts, Paul Barham, Hyung~Won Chung, Charles Sutton, Sebastian
  Gehrmann, Parker Schuh, Kensen Shi, Sasha Tsvyashchenko, Joshua Maynez,
  Abhishek Rao, Parker Barnes, Yi~Tay, Noam Shazeer, Vinodkumar Prabhakaran,
  Emily Reif, Nan Du, Ben Hutchinson, Reiner Pope, James Bradbury, Jacob
  Austin, Michael Isard, Guy Gur{-}Ari, Pengcheng Yin, Toju Duke, Anselm
  Levskaya, Sanjay Ghemawat, Sunipa Dev, Henryk Michalewski, Xavier Garcia,
  Vedant Misra, Kevin Robinson, Liam Fedus, Denny Zhou, Daphne Ippolito, David
  Luan, Hyeontaek Lim, Barret Zoph, Alexander Spiridonov, Ryan Sepassi, David
  Dohan, Shivani Agrawal, Mark Omernick, Andrew~M. Dai,
  Thanumalayan~Sankaranarayana Pillai, Marie Pellat, Aitor Lewkowycz, Erica
  Moreira, Rewon Child, Oleksandr Polozov, Katherine Lee, Zongwei Zhou, Xuezhi
  Wang, Brennan Saeta, Mark Diaz, Orhan Firat, Michele Catasta, Jason Wei,
  Kathy Meier{-}Hellstern, Douglas Eck, Jeff Dean, Slav Petrov, and Noah
  Fiedel. 2022.
\newblock Palm: Scaling language modeling with pathways.
\newblock \emph{CoRR}, abs/2204.02311.

\bibitem[{Christiano et~al.(2017)Christiano, Leike, Brown, Martic, Legg, and
  Amodei}]{RLHF}
Paul~F. Christiano, Jan Leike, Tom~B. Brown, Miljan Martic, Shane Legg, and
  Dario Amodei. 2017.
\newblock Deep reinforcement learning from human preferences.
\newblock In \emph{{NIPS}}, pages 4299--4307.

\bibitem[{Du et~al.(2022)Du, Qian, Liu, Ding, Qiu, Yang, and Tang}]{chatglm2}
Zhengxiao Du, Yujie Qian, Xiao Liu, Ming Ding, Jiezhong Qiu, Zhilin Yang, and
  Jie Tang. 2022.
\newblock Glm: General language model pretraining with autoregressive blank
  infilling.
\newblock In \emph{Proceedings of the 60th Annual Meeting of the Association
  for Computational Linguistics (Volume 1: Long Papers)}, pages 320--335.

\bibitem[{Hou et~al.(2023)Hou, Zhang, Lin, Lu, Xie, McAuley, and Zhao}]{LLMREC}
Yupeng Hou, Junjie Zhang, Zihan Lin, Hongyu Lu, Ruobing Xie, Julian McAuley,
  and Wayne~Xin Zhao. 2023.
\newblock \href {http://arxiv.org/abs/2305.08845} {Large language models are
  zero-shot rankers for recommender systems}.

\bibitem[{Jiang et~al.(2020)Jiang, Xu, Araki, and Neubig}]{world-knowledge-3}
Zhengbao Jiang, Frank~F. Xu, Jun Araki, and Graham Neubig. 2020.
\newblock \href {https://doi.org/10.1162/tacl_a_00324} {{How Can We Know What
  Language Models Know?}}
\newblock \emph{Transactions of the Association for Computational Linguistics},
  8:423--438.

\bibitem[{Jiang et~al.(2023)Jiang, Xu, Gao, Sun, Liu, Dwivedi-Yu, Yang, Callan,
  and Neubig}]{RAG2}
Zhengbao Jiang, Frank~F. Xu, Luyu Gao, Zhiqing Sun, Qian Liu, Jane Dwivedi-Yu,
  Yiming Yang, Jamie Callan, and Graham Neubig. 2023.
\newblock \href {http://arxiv.org/abs/2305.06983} {Active retrieval augmented
  generation}.

\bibitem[{Johnson et~al.(2019)Johnson, Douze, and J{\'e}gou}]{faiss}
Jeff Johnson, Matthijs Douze, and Herv{\'e} J{\'e}gou. 2019.
\newblock Billion-scale similarity search with {GPUs}.
\newblock \emph{IEEE Transactions on Big Data}, 7(3):535--547.

\bibitem[{Mao et~al.(2023)Mao, Dou, Chen, Mo, and Qian}]{LLMCDR}
Kelong Mao, Zhicheng Dou, Haonan Chen, Fengran Mo, and Hongjin Qian. 2023.
\newblock \href {http://arxiv.org/abs/2303.06573} {Large language models know
  your contextual search intent: A prompting framework for conversational
  search}.

\bibitem[{Moslem et~al.(2023)Moslem, Haque, Kelleher, and Way}]{LLMnlp}
Yasmin Moslem, Rejwanul Haque, John~D. Kelleher, and Andy Way. 2023.
\newblock \href {http://arxiv.org/abs/2301.13294} {Adaptive machine translation
  with large language models}.

\bibitem[{Nakano et~al.(2022)Nakano, Hilton, Balaji, Wu, Ouyang, Kim, Hesse,
  Jain, Kosaraju, Saunders, Jiang, Cobbe, Eloundou, Krueger, Button, Knight,
  Chess, and Schulman}]{WebGPT}
Reiichiro Nakano, Jacob Hilton, Suchir Balaji, Jeff Wu, Long Ouyang, Christina
  Kim, Christopher Hesse, Shantanu Jain, Vineet Kosaraju, William Saunders,
  Xu~Jiang, Karl Cobbe, Tyna Eloundou, Gretchen Krueger, Kevin Button, Matthew
  Knight, Benjamin Chess, and John Schulman. 2022.
\newblock \href {http://arxiv.org/abs/2112.09332} {Webgpt: Browser-assisted
  question-answering with human feedback}.

\bibitem[{OpenAI(2023)}]{GPT4}
OpenAI. 2023.
\newblock \href {http://arxiv.org/abs/2303.08774} {Gpt-4 technical report}.

\bibitem[{Ouyang et~al.(2022)Ouyang, Wu, Jiang, Almeida, Wainwright, Mishkin,
  Zhang, Agarwal, Slama, Ray, Schulman, Hilton, Kelton, Miller, Simens, Askell,
  Welinder, Christiano, Leike, and Lowe}]{instructGPT}
Long Ouyang, Jeffrey Wu, Xu~Jiang, Diogo Almeida, Carroll~L. Wainwright, Pamela
  Mishkin, Chong Zhang, Sandhini Agarwal, Katarina Slama, Alex Ray, John
  Schulman, Jacob Hilton, Fraser Kelton, Luke Miller, Maddie Simens, Amanda
  Askell, Peter Welinder, Paul~F. Christiano, Jan Leike, and Ryan Lowe. 2022.
\newblock Training language models to follow instructions with human feedback.
\newblock In \emph{NeurIPS}.

\bibitem[{Petroni et~al.(2019)Petroni, Rockt{\"a}schel, Riedel, Lewis, Bakhtin,
  Wu, and Miller}]{world-knowledge-1}
Fabio Petroni, Tim Rockt{\"a}schel, Sebastian Riedel, Patrick Lewis, Anton
  Bakhtin, Yuxiang Wu, and Alexander Miller. 2019.
\newblock \href {https://doi.org/10.18653/v1/D19-1250} {Language models as
  knowledge bases?}
\newblock In \emph{Proceedings of the 2019 Conference on Empirical Methods in
  Natural Language Processing and the 9th International Joint Conference on
  Natural Language Processing (EMNLP-IJCNLP)}, pages 2463--2473, Hong Kong,
  China. Association for Computational Linguistics.

\bibitem[{Roberts et~al.(2020)Roberts, Raffel, and Shazeer}]{world-knowledge-2}
Adam Roberts, Colin Raffel, and Noam Shazeer. 2020.
\newblock \href {https://doi.org/10.18653/v1/2020.emnlp-main.437} {How much
  knowledge can you pack into the parameters of a language model?}
\newblock In \emph{Proceedings of the 2020 Conference on Empirical Methods in
  Natural Language Processing (EMNLP)}, pages 5418--5426, Online. Association
  for Computational Linguistics.

\bibitem[{Shi et~al.(2023)Shi, Min, Yasunaga, Seo, James, Lewis, Zettlemoyer,
  and Yih}]{RAG1}
Weijia Shi, Sewon Min, Michihiro Yasunaga, Minjoon Seo, Rich James, Mike Lewis,
  Luke Zettlemoyer, and Wen{-}tau Yih. 2023.
\newblock {REPLUG:} retrieval-augmented black-box language models.
\newblock \emph{CoRR}, abs/2301.12652.

\bibitem[{Sun et~al.(2023)Sun, Yan, Ma, Ren, Yin, and Ren}]{searchGPT}
Weiwei Sun, Lingyong Yan, Xinyu Ma, Pengjie Ren, Dawei Yin, and Zhaochun Ren.
  2023.
\newblock \href {http://arxiv.org/abs/2304.09542} {Is chatgpt good at search?
  investigating large language models as re-ranking agent}.

\bibitem[{Taori et~al.(2023)Taori, Gulrajani, Zhang, Dubois, Li, Guestrin,
  Liang, and Hashimoto}]{alpaca}
Rohan Taori, Ishaan Gulrajani, Tianyi Zhang, Yann Dubois, Xuechen Li, Carlos
  Guestrin, Percy Liang, and Tatsunori~B. Hashimoto. 2023.
\newblock Stanford alpaca: An instruction-following llama model.
\newblock \url{https://github.com/tatsu-lab/stanford_alpaca}.

\bibitem[{Touvron et~al.(2023)Touvron, Lavril, Izacard, Martinet, Lachaux,
  Lacroix, Rozi{\`{e}}re, Goyal, Hambro, Azhar, Rodriguez, Joulin, Grave, and
  Lample}]{LLAMA}
Hugo Touvron, Thibaut Lavril, Gautier Izacard, Xavier Martinet, Marie{-}Anne
  Lachaux, Timoth{\'{e}}e Lacroix, Baptiste Rozi{\`{e}}re, Naman Goyal, Eric
  Hambro, Faisal Azhar, Aur{\'{e}}lien Rodriguez, Armand Joulin, Edouard Grave,
  and Guillaume Lample. 2023.
\newblock Llama: Open and efficient foundation language models.
\newblock \emph{CoRR}, abs/2302.13971.

\bibitem[{Wang et~al.(2023)Wang, Yang, and Wei}]{Query2Doc}
Liang Wang, Nan Yang, and Furu Wei. 2023.
\newblock Query2doc: Query expansion with large language models.
\newblock \emph{CoRR}, abs/2303.07678.

\bibitem[{Zeng et~al.(2022)Zeng, Liu, Du, Wang, Lai, Ding, Yang, Xu, Zheng, Xia
  et~al.}]{chatglm1}
Aohan Zeng, Xiao Liu, Zhengxiao Du, Zihan Wang, Hanyu Lai, Ming Ding, Zhuoyi
  Yang, Yifan Xu, Wendi Zheng, Xiao Xia, et~al. 2022.
\newblock Glm-130b: An open bilingual pre-trained model.
\newblock \emph{arXiv preprint arXiv:2210.02414}.

\bibitem[{Zhan et~al.(2022)Zhan, Ai, Liu, Mao, Xie, Zhang, and
  Ma}]{disentangled-retriever}
Jingtao Zhan, Qingyao Ai, Yiqun Liu, Jiaxin Mao, Xiaohui Xie, Min Zhang, and
  Shaoping Ma. 2022.
\newblock Disentangled modeling of domain and relevance for adaptable dense
  retrieval.
\newblock \emph{arXiv preprint arXiv:2208.05753}.

\bibitem[{Zhang et~al.(2023)Zhang, Xie, Hou, Zhao, Lin, and Wen}]{LLMREC-2}
Junjie Zhang, Ruobing Xie, Yupeng Hou, Wayne~Xin Zhao, Leyu Lin, and Ji{-}Rong
  Wen. 2023.
\newblock Recommendation as instruction following: {A} large language model
  empowered recommendation approach.
\newblock \emph{CoRR}, abs/2305.07001.

\bibitem[{Zhao et~al.(2023)Zhao, Zhou, Li, Tang, Wang, Hou, Min, Zhang, Zhang,
  Dong, Du, Yang, Chen, Chen, Jiang, Ren, Li, Tang, Liu, Liu, Nie, and
  Wen}]{LLMsurvey}
Wayne~Xin Zhao, Kun Zhou, Junyi Li, Tianyi Tang, Xiaolei Wang, Yupeng Hou,
  Yingqian Min, Beichen Zhang, Junjie Zhang, Zican Dong, Yifan Du, Chen Yang,
  Yushuo Chen, Zhipeng Chen, Jinhao Jiang, Ruiyang Ren, Yifan Li, Xinyu Tang,
  Zikang Liu, Peiyu Liu, Jian-Yun Nie, and Ji-Rong Wen. 2023.
\newblock \href {http://arxiv.org/abs/2303.18223} {A survey of large language
  models}.

\bibitem[{Zhou et~al.(2021)Zhou, Neubig, Gu, Diab, Guzm{\'a}n, Zettlemoyer, and
  Ghazvininejad}]{hallucinate}
Chunting Zhou, Graham Neubig, Jiatao Gu, Mona Diab, Francisco Guzm{\'a}n, Luke
  Zettlemoyer, and Marjan Ghazvininejad. 2021.
\newblock \href {https://doi.org/10.18653/v1/2021.findings-acl.120} {Detecting
  hallucinated content in conditional neural sequence generation}.
\newblock In \emph{Findings of the Association for Computational Linguistics:
  ACL-IJCNLP 2021}, pages 1393--1404, Online. Association for Computational
  Linguistics.

\bibitem[{Ziegler et~al.(2019)Ziegler, Stiennon, Wu, Brown, Radford, Amodei,
  Christiano, and Irving}]{RLHF-2}
Daniel~M. Ziegler, Nisan Stiennon, Jeffrey Wu, Tom~B. Brown, Alec Radford,
  Dario Amodei, Paul~F. Christiano, and Geoffrey Irving. 2019.
\newblock Fine-tuning language models from human preferences.
\newblock \emph{CoRR}, abs/1909.08593.

\end{thebibliography}
\bibliographystyle{acl_natbib}



\end{document}